\documentclass[twoside,11pt]{article}
\usepackage{amsmath,amsfonts,amssymb,verbatim,float,anysize,fancyvrb,graphicx,multicol,mathrsfs,mdframed}
\usepackage[utf8]{inputenc}
\usepackage[doublespacing]{setspace}
\usepackage[english]{babel}
\RequirePackage[colorlinks,citecolor=blue,urlcolor=blue]{hyperref}
\usepackage[numbers, super]{natbib}
\usepackage[usenames,dvipsnames, svgnames]{xcolor}
\usepackage{authblk}
\usepackage{multicol}
\usepackage{times}
\usepackage{graphicx}
\graphicspath{{figures/}}
\usepackage{booktabs}
\usepackage[font=small,labelfont=bf]{caption}
\usepackage{amsfonts, amsmath, amsthm, amssymb}
\usepackage{wrapfig}
\usepackage[margin=2cm]{geometry}
\definecolor{mygreen}{rgb}{0,0.6,0.5}
\definecolor{myblue}{rgb}{0,0.45,0.7}
\definecolor{myred}{rgb}{0.8,0.4,0}
\definecolor{mygray}{rgb}{.6,.6,.6}

\parskip = \baselineskip
\newlength{\normalparindent}
\AtBeginDocument{\setlength{\normalparindent}{\parindent}}

\begin{document}
\setlength\parindent{24pt}

\title{Leveraging Contact Network Information in Clustered Randomized Studies of Contagion Processes}

\author[1]{Maxwell H Wang \thanks{Maxwell Wang and Patrick Staples contributed equally to this paper.}\thanks{email: maxwang@hsph.harvard.edu}}
\author[1]{Patrick Staples$^*$\thanks{email: patrickstaples@fas.harvard.edu}}
\author[2]{M\'elanie Prague\thanks{email: melanie.prague@inria.fr}}
\author[3]{Ravi Goyal\thanks{email: r1goyal@health.ucsd.edu}}
\author[1]{Victor De Gruttola\thanks{email: degrut@hsph.harvard.edu}}
\author[1]{Jukka-Pekka Onnela\thanks{email: onnela@hsph.harvard.edu}}
\affil[1]{\small{Harvard TH Chan School of Public Health}}
\affil[2]{\small{University of Bordeaux}}
\affil[3]{\small{University of California San Diego}}

\maketitle

\begin{abstract}%
In a randomized study, leveraging covariates related to the outcome (e.g. disease status) may produce less variable estimates of the effect of exposure. For contagion processes operating on a contact network, transmission can only occur through ties that connect affected and unaffected individuals; the outcome of such a process is known to depend intimately on the structure of the network. In this paper, we investigate the use of contact network features as efficiency covariates in exposure effect estimation. Using augmented generalized estimating equations (GEE), we estimate how gains in efficiency depend on  the network structure and spread of the contagious agent or behavior. We apply this approach to simulated randomized trials using a stochastic compartmental contagion model on a collection of model-based contact networks and compare the bias, power, and variance of the estimated exposure effects using an assortment of network covariate adjustment strategies. We also demonstrate the use of network-augmented GEEs on a clustered randomized trial evaluating the effects of wastewater monitoring on COVID-19 cases in residential buildings at the the University of California San Diego.
\end{abstract}

\section{Introduction}

Contact networks capture the structure of possible pairwise transmissions (represented by network edges) in a population of actors (represented by network nodes) for various types of contagion processes; these may describe the spread of pathogens, behaviors, or ideas. Transmission on the network can only occur through edges that connect exposed and unexposed individuals; given that the structure of the network constrains pairwise transmissions, the outcome of such a process must depend on this structure. In general, the relationship between network structure and contagion processes can be complex \citep{frank1986,  newman2010book}; in investigation of exposure effects from observational data, one must consider that  network properties may confound such effects.  Furthermore, knowledge of the degree to which network features predict outcome can also  improve efficiency of estimation.

This paper investigates the question of the degree to which  incorporating information about contact network structure and summaries of contagion process outcomes at baseline improve the accuracy of estimates of exposure effects on outcomes that reflect  a contagion process operating on a network. Regardless of whether a study is observational or randomized, as long as individual outcomes  are correlated only within discrete and independent clusters but not across them, estimates of the treatment or exposure effect that ignore correlation may still be unbiased—provided that all confounding factors are measured and appropriately modeled in the observational setting. Consistent estimation of the variance must adjust for within-cluster outcome  correlation  \citep{rosenbaum2002, murray1998book, eldridge2012book}. Of note, in both randomized and observational studies, contact network information can improve statistical efficiency of estimation \citep{chemie2014}.

To accommodate the inter-cluster correlation of outcomes, we make use of Generalized Estimating Equations (GEEs) \citep{zeger1986} that provide estimates of the average marginal treatment effect across all clusters in cluster randomized trials or in settings characterized by a  correlation across groups of individuals \citep{murray1998book}. In this manuscript, we consider the augmented GEE \citep{prague2016accounting}, which allows for the use of a user-defined outcome model to improve estimation efficiency. The variance of this augmented estimate decreases asymptotically if the covariates included in the outcome model predict  the outcome. Other semiparametric approaches with potential efficiency gains have been developed, such as targeted maximum likelihood estimation (TMLE) for clustered data \citep{balzer2017new}. Although our investigation focuses on bias and efficiency using the augmented GEE, the issues we discuss are relevant for the TMLE approach as well.

The paper is structured as follows. Section 2 provides some background on networks and presents the details of the estimation procedure, including how to incorporate contact network structure in the estimation process. Section 3 demonstrates the ability of our methods to evaluate the potential gains in efficiency under different conditions in a simulation study that considers a range of network types. Section 4 considers a data example from a clustered randomized trial conducted at the University of California San Diego. Section 5 provides concluding remarks.

\section{Methods}

\subsection{Networks}

This section provides background on network concepts and the notation used throughout this paper. We assume that the true data generating mechanism is a simple susceptible-infected (SI) contagion process spreading through a network. We identify a network as a collection of nodes, such that outcomes within the network are correlated while outcomes across networks are independent. Complete data on network and contagion processes are not generally observable; but even when they are not, it may nevertheless be possible to characterize certain features of processes and/or networks. 

A simple network $\mathcal{G}$ consists of set of nodes $\mathcal{N}=\{1,...,n\}$ and edges $\mathcal{E}\subseteq \mathcal{N}\times\mathcal{N}$.  The placement of edges may be described by an $n\times n$ symmetric \emph{adjacency matrix} $\mathbf{e},$ where element $\mathbf{e}_{ij}=\mathbf{e}_{ji}$ is 1 if an edge exists between nodes $i$ and $j$ and is $0$ otherwise. The \emph{degree} of node $i$, denoted as $k_i$, is the number of edges that are adjacent to it: $k_{i}=\sum_{j}\mathbf{e}_{ij}$. Mean neighbor degree of node $i$, $\sum_{j}\frac{\mathbf{e}_{ij}k_{j}}{k_{i}}$, is the unweighted average of a node's neighbors' degrees.  \emph{Degree assortativity} is a composite measure of mean neighbor degree across the network, defined as the Pearson correlation coefficient of degrees of adjacent nodes taken over all network edges \citep{callaway2001}, which may be calculated as follows. The concept of \emph{excess degree} is often used in analytical treatment of network models; the excess degree of a node is defined as one less than the (actual) degree of the node. Let the marginal probability of a node having excess degree $k$ be $q_k$, and let the probability of this node connecting with a node with excess degree $k'$ be $p_{kk'}$. Degree assortativity can then be calculated for a given network using a sum across all degree pairs \citep{newman2002a} as $\frac{1}{{\sigma^2_q}}\sum_{kk'}kk'(p_{kk'} -q_k q_{k'})$, where $\sigma^2_q$ is the variance of the excess degree distribution for the network. A \emph{connected component} is a maximal subset of nodes for which a path exists between each pair of nodes. A \emph{path} exists between two nodes $i$ and $j$ if and only if there exists a subset of edges $\mathcal{E}_{ij}\subseteq \mathcal{E}$ in the network that connect nodes $i$ and $j$. Let $C$ be the number of components in the network, and node $i$ belongs to the connected component with label $c_{i} \in \{1,...,C\}$, where component labels are assumed to be ordered from largest to smallest. The largest connected component (LCC) contains $\sum_i\mathbb{I}(c_{i}=1)$ nodes.  The mean component size is $n/C$.

 \begin{figure}[H]
\begin{center}\vspace{0.1cm}
 \hspace*{3cm} 
\includegraphics[page=1, width=12cm, trim={3cm 3cm 3cm 2.5cm},clip]{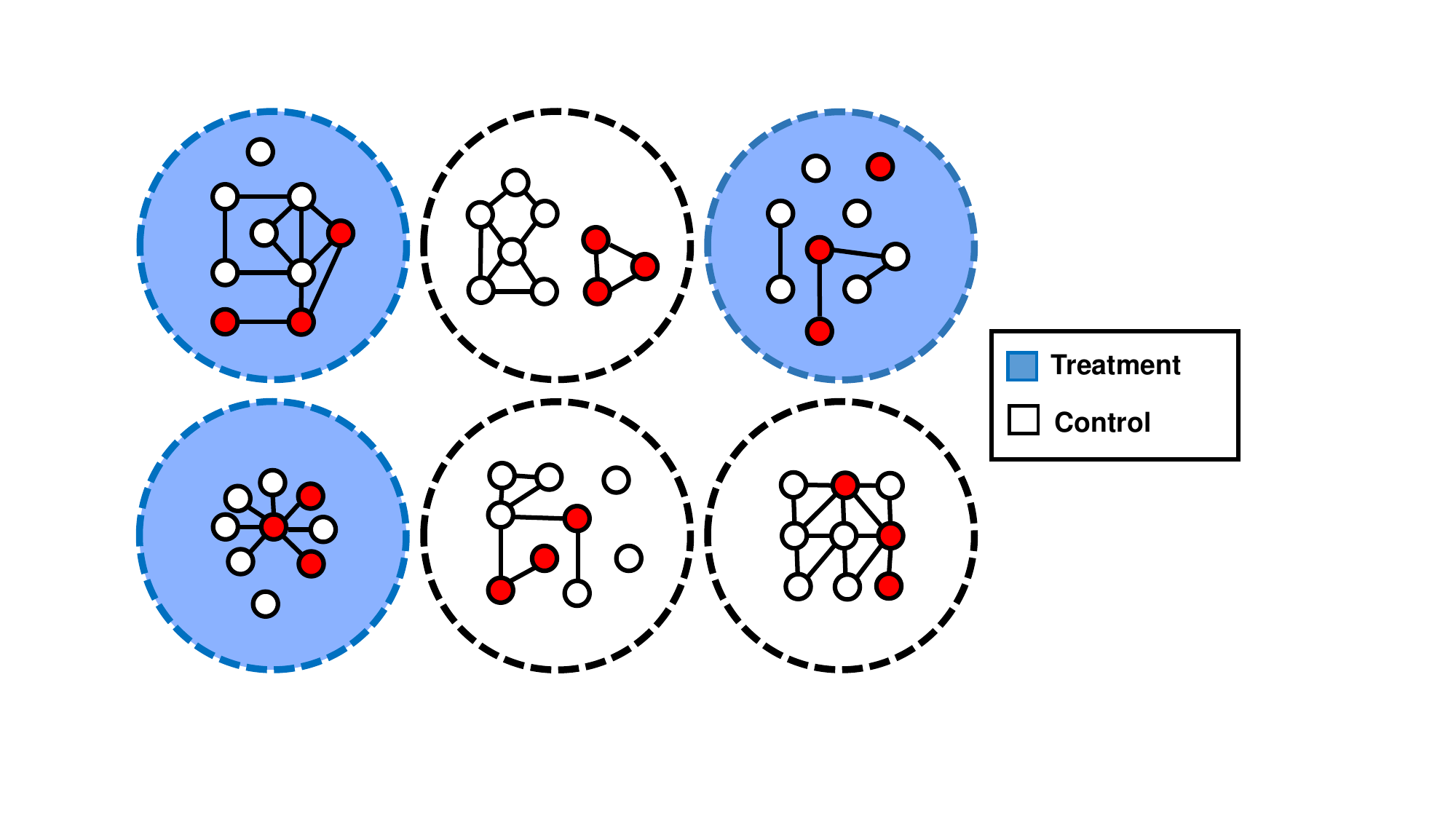}
\caption{Six clusters immediately after randomization in an idealized clustered randomized trial. Three clusters have been randomized to treatment (blue), and three have been randomized to control (white). Each cluster has nine nodes, three of which are already infected at baseline and are shaded red. Internal network structure varies between clusters, and differing cluster-level covariates can be calculated for each cluster. For example, in the fifth (bottom-center) cluster, the mean degree is 2, LCC size is 7 ($X^{(5)}$), and 3 total components are present ($X^{(8)}$). Individual outcomes in separate clusters are independent, as no edges exist between clusters.} \label{clusters}
\end{center}\vspace{0.1cm}
\end{figure}

The contagion status of each node or that of the node's network neighbors might be observable at baseline. We describe a person who has been impacted by the contagion process as affected (\emph{e.g.}, infected if the process is infectious or impacted if the process alters behavior), and we use $I_{i}(t)$ to denote the binary process outcome for node $i$ at time $t$. One simple metric of contagion status is the number of affected neighbors at baseline for node $i$, $\sum_{j}\mathbf{e}_{ij}I_{j}(0)$, or the number of affected individuals at baseline belonging to the same component as a given node $\sum_{j}\mathbb{I}(c_{i}=c_{j})I_{j}(0)$.  Another metric is the length of the shortest path between each node $i$ and each infected individual $j$ in the network at baseline.  The shortest path length between nodes $i$ and $j$ is $d_{ij}$, where $d_{ij}:=\infty$ when no path exists between the two nodes.  The shortest path length from the closest node affected at baseline is $\min_{j} d_{ij}$.  The sum of the inverse path lengths to node $i$ is $\sum_{j} \left(d_{ij}\right)^{-1}$.  Some of these metrics may be difficult to determine in practice given limited knowledge about a network or process outcomes; our interest lies in  examining whether their inclusion in  analyses yields strong enough improvements in estimation to justify the efforts necessary to gather the required data. Table \ref{feature_table} summarizes these network features.

\begin{table}[H]
	
	\begin{tabular}{ll}
		{$X^{(0)}$: Status of node $i$} at baseline & $I_{i}(0)$ \\[-.0cm]
		{$X^{(1)}$: Degree of node $i$} & $k_{i}$ \\[-.0cm]
		{$X^{(2)}$: Mean neighbor degree of node $i$} & $\sum_{j}\frac{e_{ij}k_{j}}{k_{i}}$ \\[-.0cm]
		{$X^{(3)}$: Assortativity} & See text \\[-.0cm]
		{$X^{(4)}$: Member of LCC} & $\mathbb{I}(c_{i} = 1)$ \\[-.0cm]
		{$X^{(5)}$: Size of largest component} & $\sum_i\mathbb{I}(c_{i}=1)$ \\[-.0cm]
		{$X^{(6)}$: Mean component size} & $n/C$ \\[-.0cm]
		{$X^{(7)}$: Number of components} & $C$ \\[-.0cm]
		{$X^{(8)}$: Size of node $i$'s component} & $\sum_{j}\mathbb{I}(c_{i}=c_{j})$ \\[-.0cm]
		{$X^{(9)}$: Number of neighbors affected at baseline} & $\sum_{j}\mathbf{e}_{ij}I_{j}(0)$ \\[-.0cm]
		{$X^{(10)}$:  Number of affected nodes within component at baseline} & $\sum_{j}\mathbb{I}(c_{i}=c_{j})I_{ij}(0)$ \\[-.0cm]
		{$X^{(11)}$: 1/nearest affected path length at baseline} & $\left(\min_{j} d_{ij}\right)^{-1}$ \\[-.0cm]
		{$X^{(12)}$: $\sum_{j}$ 1/path length to affected $j$ at baseline} & $\sum_{j} \left(d_{ij}\right)^{-1}$ \\[-.0cm]
	\end{tabular}

	\caption{A collection of summaries of the contact network and contagion status at baseline.}
	\label{feature_table}
\end{table}

\subsection{GEE-based estimation of the effect of a randomized exposure on outcome}

Generalized estimating equations (GEE)  provide a general approach for analyzing correlated outcomes that: i) is more robust to variance structure misspecification and relies less on parametric assumptions than the standard likelihood methods, and ii) provides population level conclusions on the effect of an exposure on an outcome.
This section reviews the augmented GEE described in \citep{prague2016accounting} and describes incorporation of contact network structure in the estimation process.

Consider an randomized study of a contagion process that consists of $k=1,...,m$ clusters with $i=1,...,n_k$ individuals per cluster, and $\sum_k n_i= n$ is the total number of individuals in the study. The binary outcome for individual $j$ in cluster $i$, $Y_{ij}$, is 1 if the individual is affected by the process by the end of the study, otherwise $Y_{ij}=0$.  $\mathbf{Y}_i=(Y_{i1}, ..., Y_{in_i})^\top$ denotes  the associated vector of outcomes in cluster $i$. We assume that there is no mixing across clusters and that outcomes are independent across them. We consider a setting where some of the clusters are randomized to a specific treatment, intervention, or exposure while others are not; we use $A_i=a$ to denote an exposure indicator such that $a=1$ for the exposed clusters and $a=0$ for the unexposed (control) clusters. In Figure \ref{clusters}, we show a schematic of such a trial. The outcome can be modeled as a function of exposure such that $ \widehat{\mathbb{E}}(\mathbf {Y}_{i})=\boldsymbol{\mu}_i(\boldsymbol{\beta},A_i)=[h^{-1}(\beta_0 + \beta_A A_{i})]_{i=1,...,n}$, with a link function $h$. The general form of a classical GEE is
$ \boldsymbol{U}(\beta) = \sum_{i=1}^m \mathbf{D}_i^T \mathbf {V}_i^{-1} \{ \mathbf {Y}_i - \boldsymbol{\mu}_i(\boldsymbol{\beta},A_i)\} \,\!$, where 
$\mathbf{D}_i=\frac{\partial \boldsymbol{\mu}_i(\boldsymbol{\beta},A_i)}{\partial \boldsymbol{\beta}^T}$ is the design matrix, $\mathbf{V}_i$ is the covariance matrix equal to $\phi \mathbf{R}_i^{1/2} \mathbf{C}(\mathbf{\alpha}) \mathbf{R}_i^{1/2}$, $\mathbf{R}_i$ a diagonal matrix with elements ${\rm var}(Y_{ij})$, $\phi$ is the dispersion parameter, and $\mathbf{C}(\boldsymbol{\alpha})$ is the ``working'' correlation structure with non-diagonal terms $\boldsymbol{\alpha}$. Parameters are estimated by setting $\boldsymbol{U}(\beta)$ to $\boldsymbol{0}$.
Because our goal is to estimate the effect of exposure, our causal parameter of interest $\beta_A$ is given as: $h^{-1}\{\mathbb{E}(Y|A=1)\}-h^{-1}\{\mathbb{E}(Y|A=0)\}$.

We assume each individual $i$ in cluster $j$ to have a set of $P$ cluster-level and individual-level covariates $\mathbf{X}_{ij}=(X^{(1)}_{ij}, ..., X^{(P)}_{ij})^\top$, with all covariates represented compactly as $\mathbf{X}_i = (\mathbf{X}_{i1}, ..., \mathbf{X}_{in_i})^\top$. Some of the covariates may relate to the network in which the individual is embedded and are described in the previous subsection. Even when the effect of intervention is not confounded, such as in a randomized setting, there exist chance imbalances in the post-randomization distribution of covariate values across treatment arms. It is then possible to improve efficiency of estimation by introducing covariate adjustment to the standard GEE framework by augmenting the GEE itself. This requires specification of an outcome model (OM) $\mathbf{B}_i(\mathbf{X}_i, A_i=a,\boldsymbol{\eta}_B)=[B_{ij}(\mathbf{X}_i,A_i=a,\boldsymbol{\eta}_B)]_{j=1,\dots,n_i}$, which is an arbitrary function of $\mathbf{X}_i$ for each exposure level, and $\boldsymbol{\eta}_B$ are nuisance parameters to be estimated. Estimation is most efficient if the OM correctly models the probability of the outcome of interest given baseline covariates $\mathbb{E}(Y_{ij}|\mathbf{X}_i,  A_i=a)$. We will call the OM ``correctly specified'' as it corresponds to the true data generation process. Equation \ref{eq:DREE} incorporates these additional terms in the estimating function. 

\begin{eqnarray}
0&=& \sum_{i=1}^m \Bigg[  \mathbf{D}_i^T \mathbf{V}_i^{-1}  \left( \mathbf{Y}_i - \boldsymbol{\mu}_i(\boldsymbol{\beta},A_i) \right)\nonumber \\ 
& &  \quad - \sum_{a=0,1}  \pi^a(1-\pi)^{1-a}\mathbf{D}_i^T \mathbf{V}_i^{-1}  \Big( \mathbf{B}_i(\mathbf{X}_i,A_i=a,\boldsymbol{\eta}_B) -\boldsymbol{\mu}_i(\boldsymbol{\beta},A_i=a)\Big) \Bigg] \label{eq:DREE}
\end{eqnarray}
The variance of $\boldsymbol{\beta}$ is estimated by using an empirical or ``sandwich'' estimator, which is also robust in the sense that it provides valid standard errors even when the assumed covariance structure is not correct. Given the OM, the exposure effect is represented by the vector of coefficients $\boldsymbol{\beta}$. 

In this paper, one can estimate the OM as $\mathbb{E}(Y_{ij}|\mathbf{X}_i,  A_i=a)$ based on some regression model, such as a simple GLM. Due to potential treatment-covariate interaction, we fit separate OM models for the two treatment arms. In our simulation study, we will consider (1) augmented GEEs adjusting for each of the thirteen covariates listed in Table 1 alone, (2) an augmented GEE adjusting for all covariates, and (3) an augmented GEE employing a stepwise selection of relevant network network covariates $\mathbf{X}_i$.  

In settings with missing data, one can include an additional propensity score (PS) term to account for data missingness. Including both the OM and the PS yields the doubly-robust (DR) GEE, which is consistent and asymptotically normal (CAN) if either the OM or PS are correctly specified \citep{prague2016accounting}. Our investigation focuses solely on the use of network covariates in the augmented GEE, but this approach can be generalized to the DR GEE for studies with missing data.

\section{Simulation Study}

In this section, we describe simulation of a contagion process on a network in a randomized study setting to estimate the effect of an intervention on that process. We use this simulation study to investigate the usefulness of our method to reduce variance of estimates of effects of exposure.  As before, we assume that individuals are nested within a collection of independent clusters, each with its own contact network structure, and that the outcome of interest arises from a contagion process propagating through network ties. We also assume that the intervention will reduce the rate of the contagion by varying amounts.

We first describe the network generation for each cluster and the contagious spreading processes as well as the effect of the exposure (or intervention) on the latter. To estimate the average exposure effect, we apply the augmented GEE described above and compare results to a standard GEE and the effect of the simulation conditions on their relative efficiencies.
\subsection*{Simulated Contact Networks}

The network generation model in our simulation study is the degree-corrected stochastic block model with degree correlation. Because of the complexity of this model, it is not possible to analytically obtain an estimate of the improvement in efficiency resulting from incorporation of network information in analyses. The original stochastic block model \citep{anderson1992_1} assumes that, in a given network, each node $i=1, \ldots, n$ belongs to only one block $b_i$ in a partition of nodes $\mathcal{B}=\{1,...,B\}$; the set of node memberships is given by the vector $\mathbf{b}=\{b_1,...,b_n\}.$  In this model, the probability of an edge between  nodes $i$ and $j$ depends only on their block membership $P(\mathbf{e}_{ij}=1)=p_{b_ib_j}.$  An extension of this model, the so-called \emph{degree-corrected stochastic block model}, allows each node $i$ to have arbitrary expected degree $\theta_i:=\mathbb{E}\left(k_{i}\right)$, where $k_{i}$ is the observed degree of node $i$ \citep{karrer2011}. The likelihood associated with this model assumes that the mean number of edges $\nu_{ij}$ between nodes $i$ and $j$ is the product of the expected degrees of nodes $i$ and $j$ ($\theta_i$ and $\theta_{j}$, respectively), multiplied by the expected amount of mixing $\omega_{b_i,b_{j}}$ between the blocks to which nodes $i$ and $j$ belong.  The full likelihood of this model is
\begin{equation}
P(\mathbf{e}\ |{\mathbf\theta}, {\mathbf\omega}, \textbf{b})=\prod_{i<j}\frac{\nu_{ij}^{\mathbf{e}_{ij}}}{\mathbf{e}_{ij}!}\exp\left(-\nu_{ij}\right)\times \prod_i\frac{\left(\frac{1}{2}\nu_{ii}\right)^{\mathbf{e}_{ij}/2}}{\left(\mathbf{e}_{ij}/2\right)!}\exp\left(-\frac{1}{2}\nu_{ii}\right), \label{DegSBMlik}
\end{equation}
where $\nu_{ij} = \theta_i \theta_{j}\omega_{b_i,b_j}$. The model assumes that $\mathbf{e}_{ij}$ is Poisson distributed, allowing for multiple edges between pairs of nodes, which converges to a simple Bernoulli network (having binary edges) for sparse networks in the limit $n\rightarrow\infty$ \citep{karrer2011}. The 1/2 terms in the second half of the likelihood account for the fact that \emph{self-edges} (edges from one node to itself) are counted twice by this indexing.

In addition to block structure and node degree, networks may vary in the extent to which degrees of adjacent nodes are correlated \citep{newman2002a}.  One metric for quantifying this property is degree assortativity, which was defined above. Degree assortativity can be varied in the network generating process by performing \emph{degree assortative rewiring}, which increases or decreases the assortativity in the network while preserving block structure and each node's degree \citep{xulvi2004,rao1996}. The details of this algorithm are given in the Appendix.

\subsubsection*{Block Structure}
\begin{figure}[H]
\begin{center}\vspace{0.1cm}
\includegraphics[page=1, width=7cm]{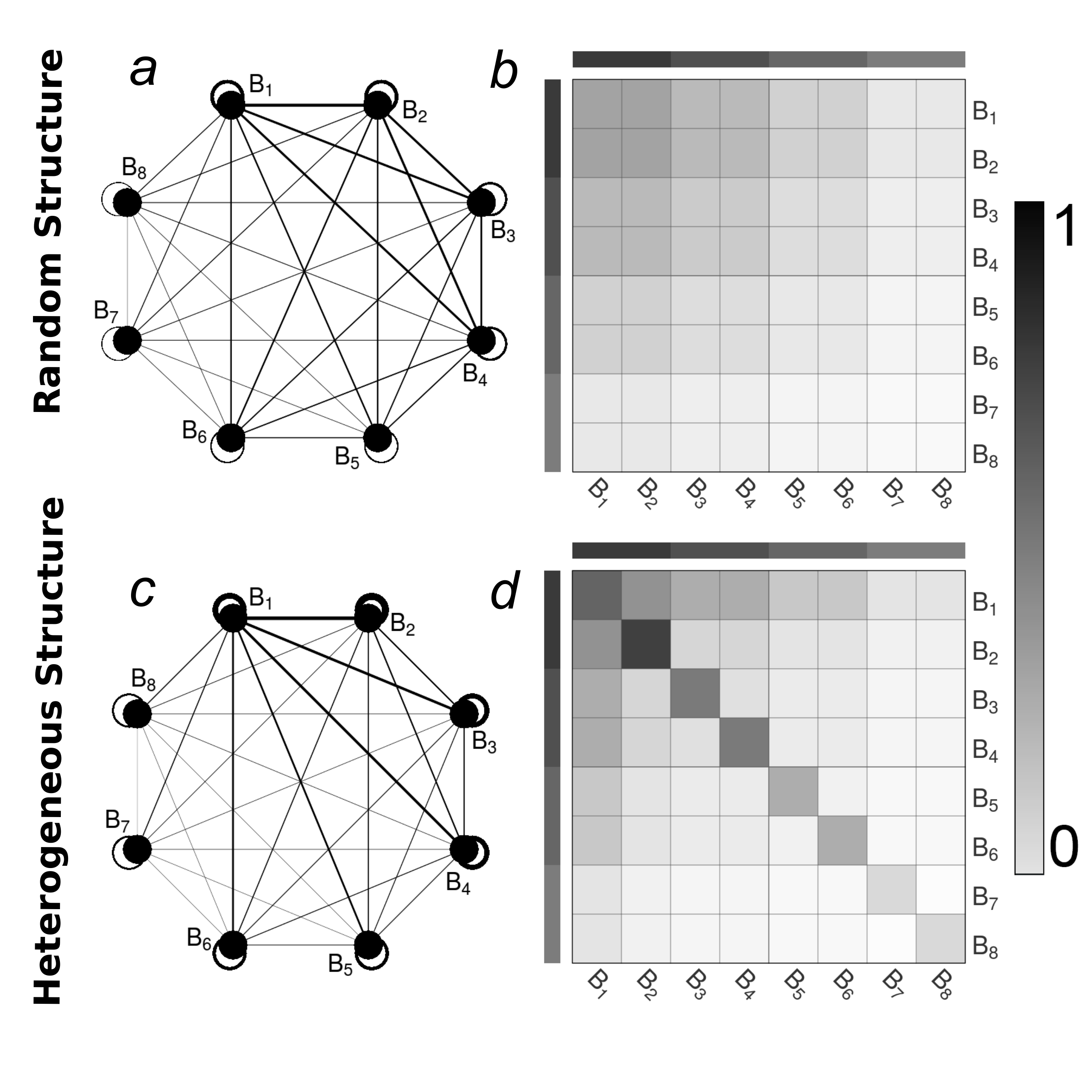}
\caption{The top and bottom rows show the two types of mixing, and the left and right panels show the block mixing structure and block mixing matrices, respectively.  In all diagrams, the eight blocks represent the eight groups of nodes that partition each network/cluster. Panel $\textbf{a}$ shows the random block structure, with lines connecting blocks that share edges.  Panel $\textbf{b}$ shows the corresponding mixing matrix, where the rows and columns represent each block, and color shade (see the color bar) represents the fraction of edges shared between members of each block. Panels $\textbf{c}$ and $\textbf{d}$ show the block mixing structure and matrices for the heterogeneous structure, respectively.} \label{block}
\end{center}\vspace{0.1cm}
\end{figure}

Each network/cluster in our simulation comprises of eight blocks. We simulate networks using two types of block structure: random and heterogeneous. For a complete description of the block structures used, see Section 2 of the Supplementary Material.  A diagram of these structures is shown in Figure \ref{block}.

\subsection*{Contagion Process}
We simulate a contagion process \citep{pastorsatorras2015}  by employing a stochastic compartmental \emph{SI} (susceptible-infectious) model \citep{anderson1992book}, shown in Algorithm 1.  Initially, $S\%$ of all nodes are selected to be affected by the contagion process (capable of transmitting) at random across all study clusters/networks $i=1,...,m$.   After initiation, affected node $j$ in network $i$ selects $z_{ij}$ of their $k_{ij}$ neighbors at random and transmits to them with probability $p_0$, where $z_{ij}$ is the node's \emph{affectivity}, which may vary between $0$ and $k_{ij}$.  Zhou \emph{et al.} showed that the properties of spreading processes on networks can depend strongly on affectivity \citep{zhou2006}.  Unit affectivity and degree affectivity occur when an individual attempts to affect either one partner (selected at random) or all partners, respectively, per unit time.  (Illustrative diagrams of the contagion process over time are given in Section 1 of the Supplementary Material.) This process is repeated until $B\%$ of the population is affected by contagion, which defines the baseline, or earliest time an infection is assumed to be observed, and can be included in the estimation procedure without risk of bias in the exposure effect estimate. Half of the clusters are exposed ($A_i=1$) and half unexposed ($A_i=0$), at random. The contagion process continues for $T$ time steps with a per-infected-node probability to infect a susceptible network neighbor $p_0$ in unexposed clusters and $p_1$ in exposed clusters. The contagion process ends at  time $T$.

\vspace{1cm}
{\small \textbf{Algorithm 1: Stochastic Compartmental Contagion Process}}
\begin{center}
\begin{itemize}
\item[1] $S\%$ of all nodes are selected uniformly at random to be initially affected.
\item[2] Until $B\%$ population incidence:
	\begin{itemize}
	\item[] For each affected node $j$ (in random order):
		\begin{itemize}
    		\item[a)] Successively select $z_{ij}$ neighbors $j_1, \mbox{...}, j_{z_{ij}}$.
    		\item[b)] If neighbor $j' \in \{j_1, \mbox{...}, j_{z_{ij}}\}$ is already affected, do nothing.  If not, affect with probability $p_0$.
		\end{itemize}
	\end{itemize}
\item[3] Repeat $T$ times:
	\begin{itemize}
	\item[] For each affected node $j$ (in random order):
		\begin{itemize}
    		\item[a)] Successively select $z_{ij}$ neighbors $j_1, \mbox{...}, j_{z_{ij}}$.
    		\item[b)] If neighbor $j' \in \{j_1, \mbox{...}, j_{z_{ij}}\}$  is already affected, do nothing.  If not, affect with probability:
    			\begin{itemize}
        		\item[] $p_0$ for those in unexposed clusters.
        		\item[] $p_1$ for those in exposed clusters.
    			\end{itemize}
		\end{itemize}
\end{itemize}
\end{itemize}
\end{center}

\subsection*{Simulation Setting Parameters}
Each simulated study consists of a contagion process propagating on  the networks associated with $m=48$ clusters of size 200, totaling 9600 individuals. The initial seed percentage $S\%$ is set to $2\%$, and the baseline discovery $B\%$ is set to $15\%$. For each scenario, we perform 1000 iterations of the simulation.

The number of clusters was chosen to be roughly comparable to cluster randomized trials investigating varied interventions such as influenza vaccines \mbox{\citep{loeb2016}} (52 clusters), hygiene programs \citep{freeman2012} (135 clusters over three treatment arms), and financial services \citep{ksoll2015} (46 clusters).The validity of the GEE robust sandwich variance estimates  requires an asymptotically large number of independent clusters. Thus, it has been suggested that GEEs only be applied when the number of clusters exceeds 30 \citep{hayes2017}. When the number of clusters is small, a variety of small-sample bias corrections can be applied \citep{mancl2001, fay2001, kauermann2001}. Another alternative are permutation tests, which may be used for valid hypothesis testing even if the number of clusters is low. For our application, we use the Fay small-sample adjustment \citep{fay2001} built into the \emph{CRTgeeDR} R package from \cite{prague2017}.

\subsubsection*{Sensitivity Analysis and Scenarios}
To estimate the sensitivity of estimation performance on simulation features, we vary four aspects of the simulation in two ways each: degree distribution (Poisson or heavy-tailed log-normal distribution), assortativity (-0.2 and 0.2), block mixing structure (random or heterogenous), and infectivity (unit or degree). This leads to a total of $2^4=16$ scenarios. 

\subsubsection*{Contagion Process}

The contagion process continues for $T=5$ time steps.  For simplicity, we assume an exposure effect that reduces contagious spread: the probability each affected node affects a selected neighbor is $p_0=0.15$ in unexposed clusters and $p_1=0.12$ in exposed clusters. We chose a small effect of intervention so that the power of the trials remains low enough to observe changes due to augmentation of the GEE. Due to major differences in transmission in scenarios including unit versus degree affectivity, we divide the probability of transmission by a constant factor of 9 in all of the scenarios exhibiting degree affectivity. This constant was chosen to attain similar levels of power between the unit affectivity and degree affectivity scenarios when estimates are obtained from the unadjusted GEE (an average of 42.6\% for the unit affectivity scenarios and 41.8\% for the degree affectivity scenarios).

\subsection*{ Evaluation of overall approach}

\subsubsection*{Performance Metrics}

In principle, the randomized setting should be free from confounding factors, and the augmented GEE approach should reduce variance in the estimation of $\beta_A$. All GEEs used an exchangeable correlation structure. Based on the simulation specifications, we define estimation metrics for the average exposure effect $\beta_A$. For $r=1,...,R$ replicates, the estimate of exposure effect is denoted $\widehat{\beta}_r$ and the estimated standard deviation $\widehat{\text{sd}}(\widehat{\beta}_r)$. These standard deviation estimates are  compared with the empirical bootstrap estimates, which are calculated  as the standard deviations of all  estimates $\ \widehat{\beta}_r$ in the R replicates. Empirical power and coverage are derived from these simulation study point estimates and their confidence intervals.  Power is based on 0.05 level tests of the exposure effect. Coverage is defined as the percentage of  replicates for which the true value of exposure effect is included in the 95\% confidence interval of its estimate. Since the contagious spreading rates in unexposed and exposed groups are specified as $p_0$ and $p_1$ at the node level, the true value for the average exposure effect must be estimated through simulation \citep{woodward2001}. To estimate the true underlying exposure effect $\beta_A$, we simulate an additional 20,000 clustered randomized trials, each with a uniquely generated network satisfying the scenario of interest, and obtain the mean treatment effect, averaged over all trials. We define improvement in estimation efficiency as the percent reduction in root mean squared error (RMSE) for each covariate set in the outcome model, comparing the augmentation adjustment $\widehat{\text{RMSE}}_\text{adj}$ to that of the unadjusted GEE ($\widehat{\text{RMSE}}_\text{GEE}$):

\begin{align}
\widehat{\text{Improvement}} := 100\times \left(1-\frac{\widehat{\text{RMSE}}_\text{adj}}{\widehat{\text{RMSE}}_\text{GEE}}\right).
\end{align}

\subsection*{Simulation Results}
Due to inherent randomness, the covariates included in the outcome model may be correlated with the exposure. By augmenting the GEE with such covariates, we can adjust for that imbalance and obtain higher efficiency. Bias in estimation, average model standard error, empirical standard error, RMSE reduction, power, and coverage are provided in Table \ref{marginal_table}; full simulation results are given in Section 3 in Supplementary Material. Results are averaged across the 16 observational scenarios with the standard deviations across scenarios shown in parentheses. Averaged across all simulation replications and variants, inclusion of covariates in outcome and propensity score models led to gains in efficiency. We also find that a single covariate, the number of affected nodes in the same component as the node at baseline ($X^{(10)}$), provides a reduction in RMSE comparable to or higher than that achieved by the variable selection approach. 

\begin{table}[H]
\begin{center}
\small
\singlespacing
\begin{tabular}{r|rrrrrr}
  & None & $X^{(0)}$ & $X^{(9)}$ & $X^{(10)}$  & All & Stepwise\\
  \hline
  Bias & 0.000 & -0.000 & -0.000 & -0.000 & -0.001 & -0.000 \\ 
  Estimated SE & 0.066 & 0.042 & 0.043 & 0.036 & 0.035 & 0.035 \\ 
  Empirical SE & 0.068 & 0.043 & 0.044 & 0.039 & 0.039 & 0.038\\ 
  Improvement (\%) & 0(0) & 36(2) & 34(10) & 43(3) & 41(5) & 43(5) \\ 
  Power & 42(5) & 77(5) & 76(11) & 87(4) & 88(4) & 89(4) \\ 
  Coverage & 94(1) & 94(1) & 94(1) & 93(1) & 92(1) & 92(1) \\
\end{tabular}
\end{center}
\caption{Exposure effect statistics averaged across each simulation characteristic, adjusted for all confounding factors. Each row displays a metric, and each column displays an adjustment feature or strategy for the outcome model. Standard deviations across scenarios are shown in parentheses. The ``None" column corresponds to a standard GEE.}
\label{marginal_table}
\end{table}

\subsubsection*{Network feature selection}
\begin{figure}[H]
\centering
\includegraphics[page=1,width=14cm,clip=true]{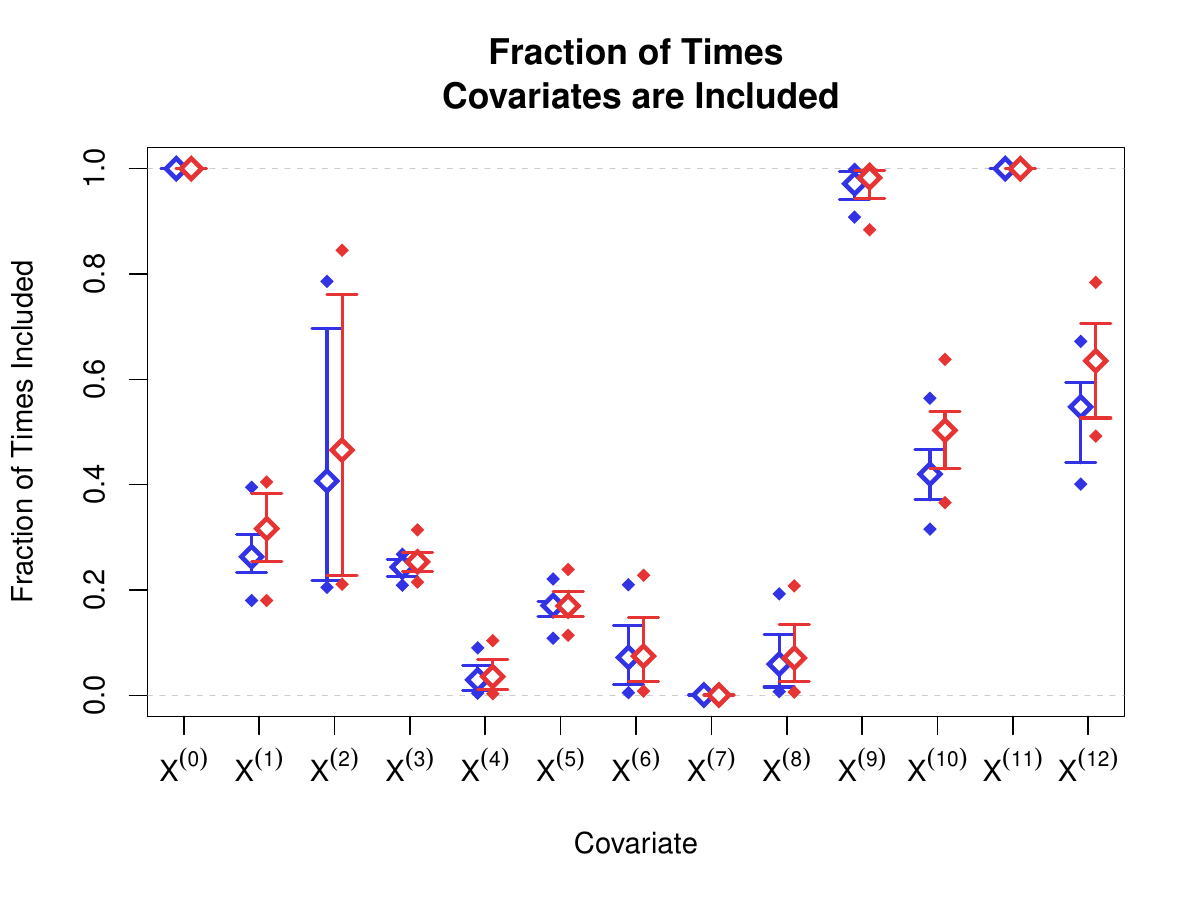}
\caption{The distribution of the proportion of occasions each covariate is included in the outcome model, under a stepwise procedure.  The covariates along the $x$-axis are described in Table \ref{feature_table}.  Blue bars show results for the exposed group, and red bars show results for the unexposed group.  The middle values represent medians, the bars represent the (25, 75)\% quartiles across scenarios, and endpoints indicate minima and maxima.}
\label{percent_included}
\end{figure}

The covariates selected in a stepwise procedure for inclusion in the outcome model vary across the different simulated datasets. To assess which covariates are most useful for adjustment in the outcome model, we measure the frequency of covariate inclusion and its variability by simulation scenario (see Figure \ref{percent_included}).  Baseline status ($X^{(0)}$), degree ($X^{(1)}$) and covariates related to contagion at baseline (covariates $X^{(9)}$ and $X^{(11)}$) are included most often; others are selected in a range of frequencies. 

\subsubsection*{Sensitivity Analysis Results}
Features of  the simulated contagion process affect performance metrics such as power and improvement in RMSE. To evaluate the sizes of these effects, we used simple linear regression treating the RMSE as the outcome and simulation features (i.e., mean degree, degree distribution, assortativity, block mixing structure, infectivity mode, and infection prevalence at baseline) as covariates; these are coded as binary variables.  The fitted coefficients represent the metric change when changing simulation features, holding all other simulation features constant. The percent improvement in RMSE is shown in Table \ref{improvement_modification_table}. For example, holding all other simulation features constant and using stepwise model selection, a contagion process exhibiting degree affectivity shows an additional RMSE reduction of 7.2 percentage points compared to a contagion process exhibiting unit affectivity.  Using covariate $X^{(9)}$ as a predictor yields much larger reductions in RMSE if the contagion exhibits degree affectivity instead of unit affectivity.  This suggests that the contagion status of a node's neighbors can be quite predictive of the risk of the node becoming affected, especially in the degree affectivity case. Table \ref{power_modification_table} shows analogous results for changes in power, holding all other simulation features constant.

\begin{table}[H]
\begin{center}
\small
\singlespacing
\begin{tabular}{r|rrrrr}\\
 &  $X^{(0)}$ & $X^{(9)}$ & $X^{(10)}$ & All & Step \\
  \hline
  (Intercept) & 35.0 & 25.4 & 39.2 & 33.1 & 37.0 \\ 
  Degree vs. Unit Affectivity & -1.4 & 19.1 & 3.8 & 8.2 & 7.2 \\
  Log-Normal vs. Poisson & -0.5 & -2.1 & -0.9 & 2.0 & -0.3 \\ 
  Assortative vs. Disassortative & 1.4 & -0.1 & 2.6 & 2.1 & 2.3 \\ 
  Communities vs. No Communities & 2.0 & 0.2 & 1.3 & 4.3 & 3.9 \\ 
\end{tabular}
\caption{Percent improvement in RMSE reduction when changing simulation assumptions.  Rows display a simulation assumption, and columns display an adjustment feature or strategy. The Intercept value represents the RMSE reduction associated with each augmentation term, under the scenario exhibiting unit affectivity, Poisson degree distribution, disassortativity, and no community structure.}
\label{improvement_modification_table}
\end{center}
\end{table}

\begin{table}[H]
\begin{center}
\small
\singlespacing
\begin{tabular}{r|rrrrrr}\\
 &  None & $X^{(0)}$ &$X^{(9)}$ & $X^{(10)}$ & All & Step \\
  \hline
  (Intercept) & 48.5 & 84.1 & 73.0 & 88.6 & 87.1 & 88.2 \\ 
  Degree vs. Unit Infectivity & -0.9 & -3.0 & 19.2 & 3.3 & 6.2 & 6.1 \\
  Log-Normal vs. Poisson & 0.4 & 0.5 & -0.2 & 0.4 & 1.6 & 0.6 \\ 
  Assortative vs. Disassortative & -2.7 & -1.7 & -2.9 & -0.1 & -1.0 & 0.5 \\ 
  Communities vs. No Communities & -9.4 & -9.1 & -10.0 & -7.0 & -4.2 & -4.3 \\ 
\end{tabular}
\caption{Percent increase in power when changing simulation assumptions.  Rows display a simulation assumption, and columns display an adjustment feature or strategy. The Intercept value represents change in power associated with each augmentation term, under the scenario exhibiting unit affectivity, Poisson degree distribution, disassortativity, and no community structure.}
\label{power_modification_table}
\end{center}
\end{table}

\section{Application to Clustered Randomized Trial}

In the following section, we demonstrate the application of our method to a CRT conducted at the University of California San Diego (UCSD).

\subsection*{Wastewater Monitoring Cluster Randomized Trial}

Wastewater monitoring paired with automated reporting systems can be utilized for forecasting COVID-19 cases and preventing outbreaks. Here, we consider the clustered randomized trial component of the wastewater surveillance program implemented at UCSD \citep{karthikeyan2021}. In this program, wastewater monitors were placed inside manholes associated with selected UCSD residential buildings. 
Between November 23 and December 29, 2020, manholes were randomized to either receive wastewater monitors or not. The intervention was also paired with an alert system that notified residents when COVID-19 was detected in the wastewater associated with their building.

In keeping with the structure of our simulation example, we define each cluster by a unique manhole. Thus, a single cluster is composed of the set of residence buildings associated with a single manhole. Each residence building is represented by a single node within the contact network associated with its respective cluster. For this simple example, we considered each cluster to be a complete (maximum density) network, such that there exists an edge between every pair of buildings that reside in the same cluster. In total, we considered 41 manholes, and the sizes of manhole-associated clusters ranged between 1 and 7 residence buildings.

During the period of this study, all on-campus UCSD students were mandated to adhere to a biweekly testing schedule. We define the outcome of interest as the total number of positive COVID-19 tests registered during the period of randomization. During the period of randomization, a total of 11930 tests were returned, with 68 of these tests indicating a positive COVID-19 result.

\subsection*{Results}

In Table \ref{marginal_table_ucsd}, we summarize results obtained by augmenting a GEE with a selection of the various network features defined in Section 2. The definition of several network covariates requires knowledge about the number of cases at baseline, prior to the delivery of randomized treatment. As a proxy for this measure, we consider the number of positive cases registered from each residence during the three months prior to the introduction of wastewater monitoring devices. During this initial baseline period, a total of 20655 tests were returned, and 28 showed a positive COVID-19 result.

In addition, we note that because the networks for every cluster was complete, some of the network features are identical. A few examples are $X^{(9)}$ and $X^{(10)}$, or $X^{(1)}$ and $X^{(2)}$. We also define an additional network covariate, $X^{(13)}$, which is defined as the sum of the total cases occurring in neighbors at baseline.

\begin{table}[H]
\begin{center}
\small
\singlespacing
\begin{tabular}{r|rrr}
  & $\beta$ & Variance & Improvement\\
  \hline
  Unaugmented & -0.691 & 0.286 & NA\\ 
  $X^{(0)}$ & -0.648 & 0.208 & 27.4\%\\
  $X^{(1)}$ & -0.618 & 0.196 & 31.5\%\\
  $X^{(9)}$ &-0.496 & 0.238 & 16.7\%\\
  $X^{(13)}$ & -0.609 & 0.215 & 24.9\%\\
\end{tabular}
\end{center}
\caption{Point estimates, robust sandwich variances, and percent improvements for selected network covariates. Not all network covariates listed in Table 1 were included, as many are redundant when applied to complete networks.}
\label{marginal_table_ucsd}
\end{table}

All of the selected network covariates led to significant improvements in variance estimates. Particularly of note is the network covariate $X^{(1)}$, the individual-level covariate of node degree. In most of our simulated scenarios, $X^{(1)}$ led only to minor improvements in efficiency, but in this data example, its associated improvement is significant. This is likely due to a higher variance in degree distribution among clusters in the UCSD trial when compared to our simulations, where degree distribution was controlled by scenario.

\section{Discussion}

Spread of a contagious process  depends on contact network structure and contagious  process dynamics.  To estimate the marginal effect of exposure on this process, augmented GEE methods as described above can be used to reduce the variation of the exposure effect estimate. Even after randomization, chance imbalances in covariates can still exist between treatment arms. Through a user-specified outcome model, estimation efficiency can be improved by taking these covariate imbalances into account \citep{stephens2012}. 

In settings where network information is available, augmenting the GEE procedure is particularly important. The space of possible networks is high-dimensional; hence network structure, even when described by summary statistics, can vary widely between clusters, and imbalance between treatment arms is almost guaranteed. Furthermore, the degree to which differing networks impact individual outcomes also varies. Due to high potential variability between treatment arms, the inclusion of relevant network covariates in the augmented GEE may lead to significant efficiency gains. 

Our focus was  investigation  of the extent to which power and RMSE in exposure effect estimates depend on specific network properties. Adjusting for contact network features and baseline contagion improved power and yielded a considerable reduction in RMSE  across a range of simulation settings.

In our simulation studies, covariates derived from a variety of network features differed in their usefulness in reducing exposure estimate variance. They also differed in the degree to which  collection of the required information would be feasible. Obtaining the degree of an individual might be relatively easy to collect in some settings by using a simple survey, but the gains we observed from using this covariate were very modest. We found that the number of nodes in the component affected at baseline yielded the largest reduction in variance; though perhaps more challenging to obtain, this quantity may also  be feasible to estimate in some important settings, such as infectious disease outbreaks. The number of neighbors affected at baseline also yielded large reductions in variance and may be estimated under similar settings. While not a network covariate, individual baseline status was selected for in the stepwise models 100\% of the time, and as a lone augmentation term, still yielded significant reductions in variance.

In addition, we applied our method to a real data example from the a clustered randomized trial on the efficacy of wastewater monitoring at UCSD. Again, augmentation using a variety of network features led to significant increases in statistical efficiency. Notably, one feature ($X^{(1)}$) that did not lead to significant efficiency gains in the simulated scenarios led to a large improvement in the data example. While the degree distribution across clusters was relatively homogeneous in the simulated scenarios, the complete networks in the data example had high variance in degree distributions. Thus, correcting for chance imbalances in degree distributions across the treatment and control arms was far more effective in the data example. For the data example, we chose to assume complete networks existed within each cluster, even though one might imagine more advanced methods of inferring contact networks between residence buildings. However, we note that augmentation with network features will lead to efficiency gains as long as the network features are imbalanced across treatment arms and are also associated with the outcome of interest. Under these conditions, the use of estimated contact networks or proxies for network features can still increase efficiency.

This work invites several extensions. Information may be missing or misreported for individual outcomes, contact network data, or both;  incompleteness of data may lead to bias or increased variation in estimating the exposure effect. Methods for addressing missing information on both networks and outcomes need to be developed. Although we carried out a wide range of simulations, the range of possible scenarios in which the methods might be relevant is large and beyond the scope of any single paper to address. Simulation of other settings would be useful to help guide randomized studies in which augmentation by network features may help improve efficiency of estimation. Appropriate methods still need to be developed for observational studies with network-related confounding. Lastly, the SI contagion model presented in this work most closely resembles the effect of an educational intervention; to model an infectious disease, SIR, SIS, and other compartmental models may be more realistic.

\section{Appendix}

\begin{figure}[H]
\includegraphics[width=8cm]{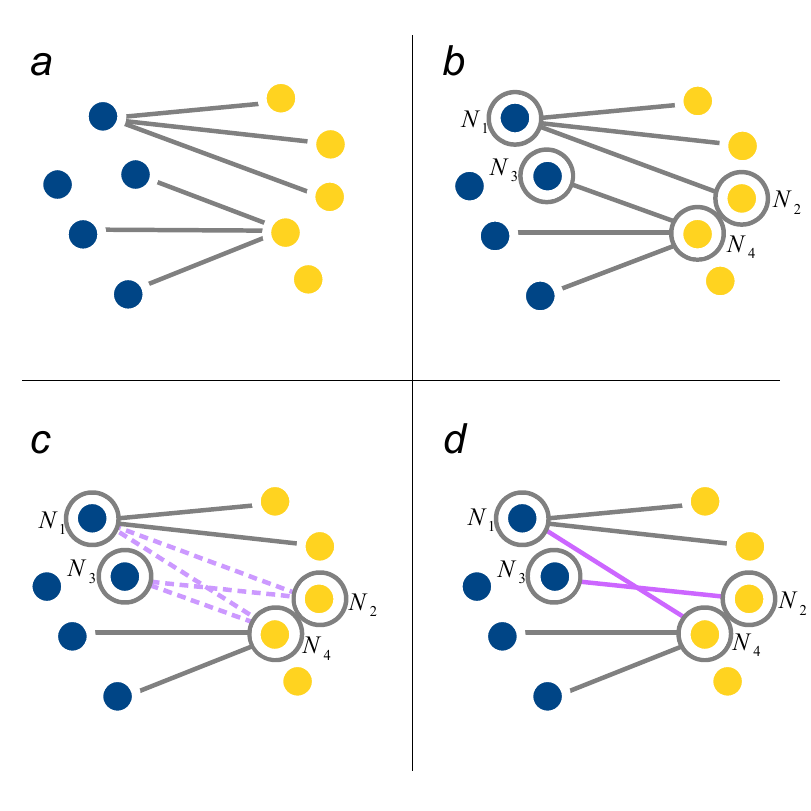}
\caption{A schematic of degree assortative rewiring.  Panel \textbf{a} displays a network, containing nodes $N_1,...,N_4$.  Panel \textbf{b} highlights two edges selected within the same block pair.  Panel \textbf{c} shows a potential rewiring, which will only occur if rewiring will increase assortativity.  In this case, rewiring would increase degree assortativity, and panel \textbf{d} displays the rewiring.}
\label{rewire}
\end{figure}

\textbf{Degree assortative rewiring.} This is performed by randomly selecting two edges within a block pair and rewiring them, as described in Algorithm 2.  A diagram of this process is shown in Figure \ref{rewire}.  To decrease assortativity, the inequality in Step 3 must be reversed. There exist cases where it is not possible to obtain high (or low) enough assortativity after specifying the degree of each node. When this occurs, we accept the network if assortativity converges to within 10\% of the desired assortativity, and regenerate the entire network if it does not.

\newpage
\vspace{.5cm}
{\small \textbf{Algorithm 2 Edge Rewiring to Adjust Assortativity}}
\begin{center}
\begin{itemize}
  \item[1] Select two blocks $b_l$ and $b_{l'}$ at random.
  \item[2] Select two edges $(N_1,N_2)$ and $(N_3,N_4)$ at random between blocks $b_{l}$ and $b_{l'}$.
  \item[3] If $|k_{N_1} - k_{N_2}| + |k_{N_3} - k_{N_4}| > |k_{N_1} - k_{N_4}| + |k_{N_2} - k_{N_3}|$:
  \item[]\hspace{1cm} Remove edges $(N_1,N_2)$ and $(N_3,N_4)$
  \item[]\hspace{1cm} Add edges $(N_1,N_4)$ and $(N_2,N_3)$
\end{itemize}
\end{center}

\section*{Acknowledgements}{
 We are grateful for support from NIH Grants T32AI007358-26, R37 AI051164, R24 AI106039, R01 AI-147441, and R01 AI138901. We thank Natasha Martin, Jingjing Zou, and Tuo Lin for their assistance in preparing the UCSD wastewater monitoring dataset as well as for their important insights about the nature and conduct of the study. We would also like to acknowledge the large number of individuals responsible for the funding, installation, and monitoring of the wastewater samplers.
}

\section*{Ethics}{
The UCSD institutional review board deemed investigation of effects of wastewater testing not to be human subject research, as personally identifiable information was not recorded.
}

\bibliography{citations}

\end{document}